\documentclass[a4paper,11pt]{article}
\usepackage{pos}
\usepackage{comment}
\usepackage[capitalize]{cleveref}
\usepackage{subcaption}  % Add this in your preamble
\usepackage{lineno}
\usepackage{enumitem}
\usepackage{wrapfig}
\usepackage{xspace}
\def\Offline{\mbox{$\overline{\textrm{Off}}$\hspace{.05em}\protect\raisebox{.4ex}{$\protect\underline{\textrm{line}}$}}\xspace}
\def\avg#1{\langle{#1}\rangle}

\title{Muon Signal Charge in the Underground Muon Detector of AugerPrime}
\ShortTitle{Muon Signal Charge in the Underground Muon Detector of AugerPrime}

\author*[a,b]{Marina Scornavacche}
  
\affiliation[a]{Instituto de Tecnolog{\'\i}as en Detecci\'on y Astropart\'iculas,\\ 
Av.\ Gral.\ Paz 1499, Buenos Aires, Argentina}

\affiliation[b]{Institut f\"ur Astroteilchenphysik, Karlsruher Institut f\"ur Technologie,\\
Hermann-von-Helmholtz-Platz 1 76344, Eggenstein-Leopoldshafen, Germany}

\onbehalf{for the Pierre Auger Collaboration$^c$}
\affiliation[c]{Observatorio Pierre Auger,\\
Av.\ San Mart{\'\i}n Norte 304, Malarg\"ue, Argentina\\
Full author list: {\rm\url{https://www.auger.org/archive/authors_icrc_2025.html}}}

\emailAdd{spokespersons@auger.org}

\abstract{

The Underground Muon Detector of the Pierre Auger Observatory features a calorimetric detection mode that estimates the number of muons based on signal charge measurements. This contribution provides an overview of the calibration procedure, revisiting the previously published strategy and identifying a bias introduced by the triggers used to estimate the mean charge deposited by a vertical muon. We demonstrate that calibrating underground detectors requires careful consideration of the interactions of penetrating particles through matter. In devices based on plastic scintillators, energy deposition—and thus the recorded charge—is significantly affected by knock-on electron production in the surrounding ground as muons traverse the medium. To mitigate this effect, we propose a new calibration strategy that ensures an unbiased muon estimator. This approach is applied to data collected from 2019 to 2024. 

}

\ConferenceLogo{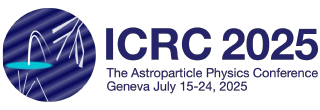}

\FullConference{39th International Cosmic Ray Conference (ICRC2025)\\
 15–24 July 2025\\
Geneva, Switzerland\\}

\begin{document}
\maketitle

\section{Introduction} \label{Introduction}

The Pierre Auger Observatory, the largest observatory for studying ultra-high-energy cosmic rays, is undergoing an upgrade, \textit{AugerPrime}~\cite{upgrade}, aimed at improving the separation between electromagnetic and muonic components of air showers produced by cosmic rays. The Underground Muon Detector (UMD) plays a crucial role in \textit{AugerPrime}, as it enables a direct measurement of the muonic component, a key observable for determining the cosmic-ray mass composition. 

\begin{figure}
\centering
    \includegraphics[width=0.43\textwidth]{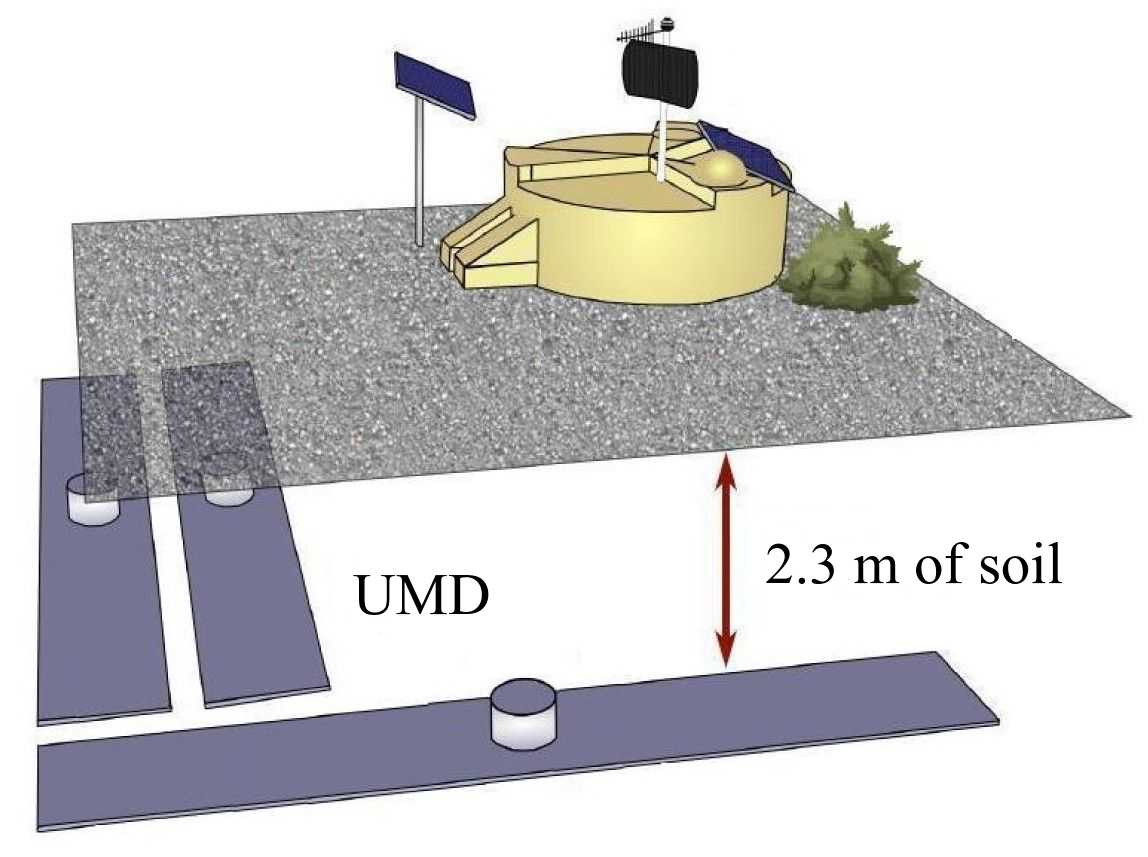}
    \hfill
    \includegraphics[width=0.5\textwidth]{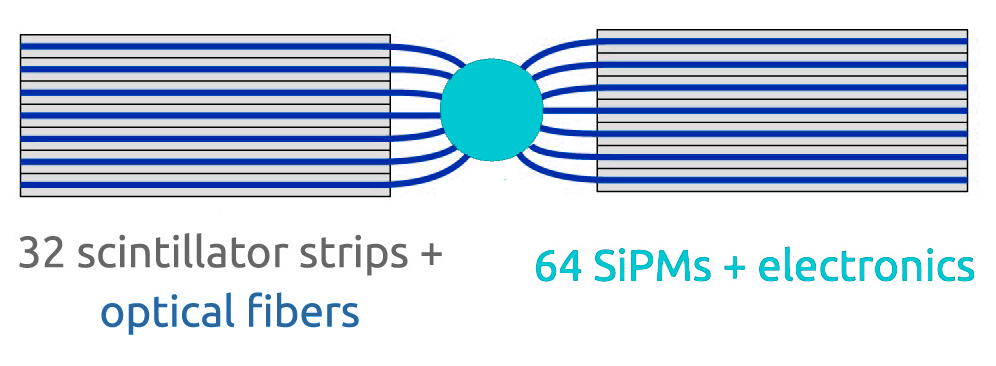}
    \caption{Schematics of the UMD design.}
    \label{DetectorDescription}
\end{figure}

The Pierre Auger Observatory consists of a 3000\,km$^2$ Surface Detector (SD), equipped with 1660  water-Cherenkov detectors distributed on a triangular grid with 1500\,m spacing, along with two denser sub-arrays with spacings of 750\,m and 433\,m covering 23.5\,km$^2$ and 1.9\,km$^2$, respectively. These two denser arrays will be equipped with 219 UMD scintillator modules, of which 61\% are currently operational, with full deployment expected by 
the end of 2025. As shown in \cref{DetectorDescription}, three UMD modules of 10 m$^{2}$ are buried at 2.3\,m next to a water-Cherenkov detector. Each module consists of 64 plastic scintillator strips of 
$400 \times 4 \times 1$\,cm containing wavelength-shifting optical fiber connected to an array of 64 silicon photomultipliers (SiPMs). When a muon impinges on the scintillator, the produced photons are collected and propagated along the fibers to the photodetector~\cite{umddesign1}.

The UMD operates in two complementary modes: binary and ADC, the latter being the calorimetric acquisition mode of the UMD. The binary mode offers better resolution at low particle densities (far from the shower core), while the ADC mode performs better at high densities (close to the shower core). The former relies on the time-over-threshold of the signals from individual SiPMs, processing each of the 64 channels independently via a discriminator~\cite{flavia, joaquincorner}; the latter integrates the charge from the summed signals of all 64 SiPMs, digitised through high- and low-gain amplifiers into 1024-sample waveforms at 6.25\,ns intervals. In this work, the low-gain channel of the ADC mode is used to estimate the muon content as follows:

\begin{equation}
N_\upmu^\text{ADC} =
  \frac{q_\text{meas}\cos\theta}{\avg{q_{1\upmu}(\theta{=}0^\circ)}},
\label{ecuacioncarga}
\end{equation}
where the quantity in the numerator is the total charge per vertical path length, with $q_\text{meas}$ the total measured charge in a 10\,m$^2$ UMD module and $\theta$ the shower zenith angle, used to approximate the mean muon zenith angle. The quantity in the denominator, $\avg{q_{1\upmu}(\theta{=}0^\circ)}$, is the average charge deposited by individual vertical muons. 

Reference~\cite{calibrationADCpaper} proposed a calibration method to estimate the mean charge deposited by vertical muons using atmospheric muons detected when a UMD module receives a local trigger from its associated water-Cherenkov detector. The algorithm identifies single-muon signatures in the binary mode to build calibration histograms. However, the trigger condition introduces a bias toward inclined, higher-energy muons, leading to asymmetries between module halves. Although the atmospheric flux is dominated by near-vertical muons, the rate at which they trigger a coincidence with the water-Cherenkov detector is an order of magnitude lower than the rate for single inclined muons that traverse both detectors. As a result, the method is not a reliable estimator of the charge deposited by vertical muons.

In this work, we show that even when the charge deposited by a single vertical muon is correctly measured, the previous calibration method remains inadequate. In \cref{seccionknockon}, we show that calibrating the ADC mode using the average charge from single vertical muons, as proposed by \cref{ecuacioncarga}, introduces a bias in the reconstructed muon number due to an increased energy deposition per muon, especially near the shower axis, where higher-energy muons produce more knock-on electrons. In \cref{seccionimproved}, we present a calibration method based on simulations that accounts for the variations in the charge deposited with the energy of the incoming muons. The same methodology is applied to data using the observables available in measured events.

\section{The effect of knock-on electrons}\label{seccionknockon}

In an inelastic collision with atoms, an energetic charged particle, such as a muon, transfers energy to the electrons bound in orbitals. When the energy transferred to the electron is higher than its ionization energy, the electron is ejected from its atomic orbital. When the ejected electron has sufficient kinetic energy to travel a significant distance from its point of ejection~\cite{knockonelectrons}, it is referred to as a knock-on electron. The higher the energy of the muon, the greater is the average energy transferred to the knock-on electrons, enabling them to traverse longer distances and enhancing their chances to reach the underground scintillators. The main source of bias in the ADC mode arises from the energy deposited by such electrons generated from the interaction of muons with the surrounding soil of the UMD~\cite{UHECRKnockOn}. A detailed analysis of these effects and their impact on muon reconstruction is provided in Ref.~\cite{UHECRKnockOn}, and only a brief summary is presented here.

Proton showers were simulated with the hadronic interaction model EPOS-LHC at energies of $10^{17.5}$, $10^{18}$, and $10^{18.5}$\,eV and zenith angles of 0, 12, 22, 32, and 38 degrees. The secondary particles of the shower that reached ground level were propagated through the soil and the energy deposition in the UMD was calculated using Geant4~\cite{Geant4}. The response of the detector was simulated in \Offline~\cite{offline}, the official software of the Pierre Auger Observatory. Saturated modules were excluded.

\begin{figure}[htbp]
\def\w{0.49}
    \centering
        \begin{subfigure}[b]{\w\textwidth}
        \includegraphics[width=\textwidth]{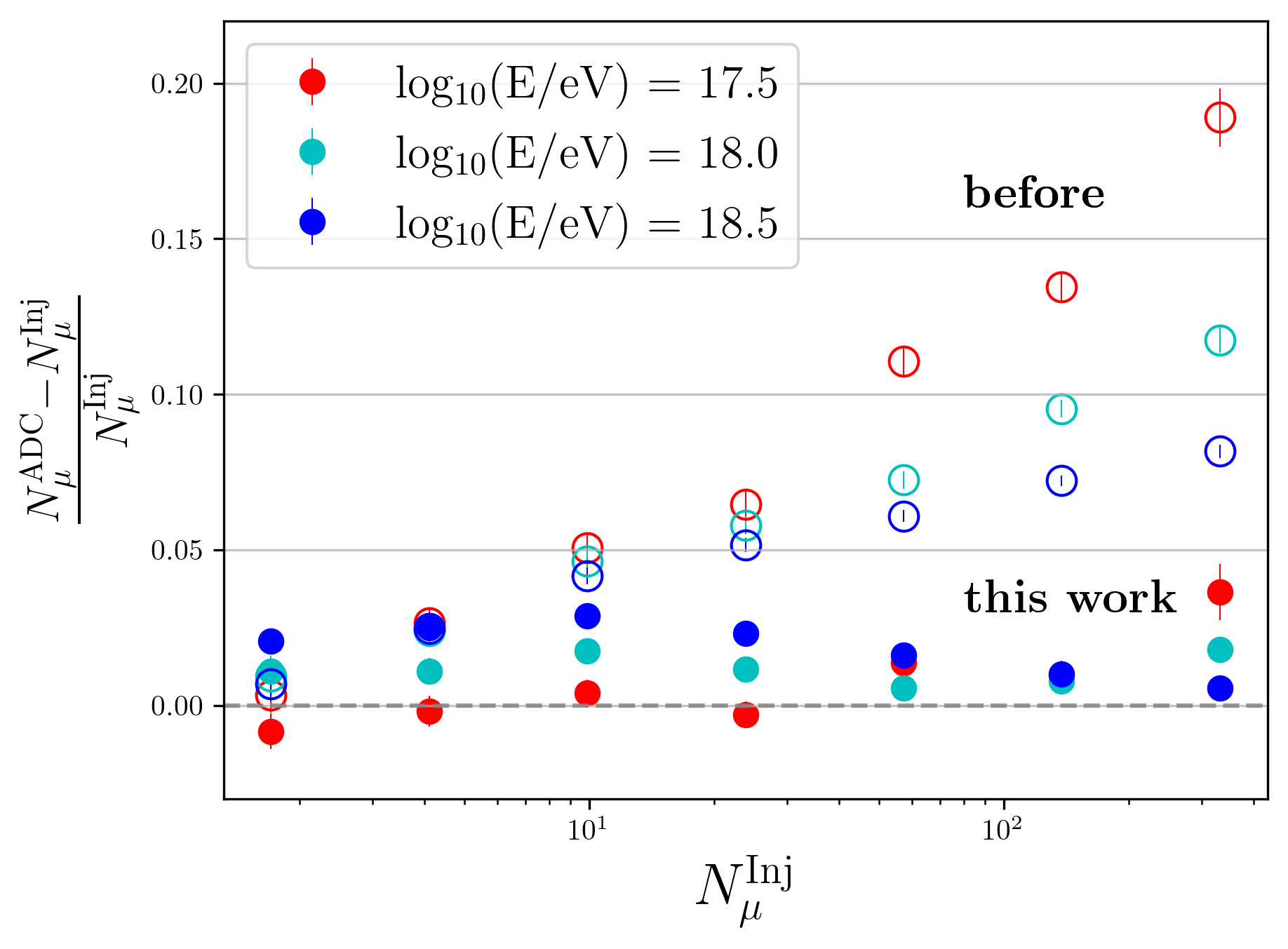}
        \caption{}
        \label{calibration_biasesb}
    \end{subfigure}
    \hfill
    \begin{subfigure}[b]{\w\textwidth}
        \includegraphics[width=\textwidth]{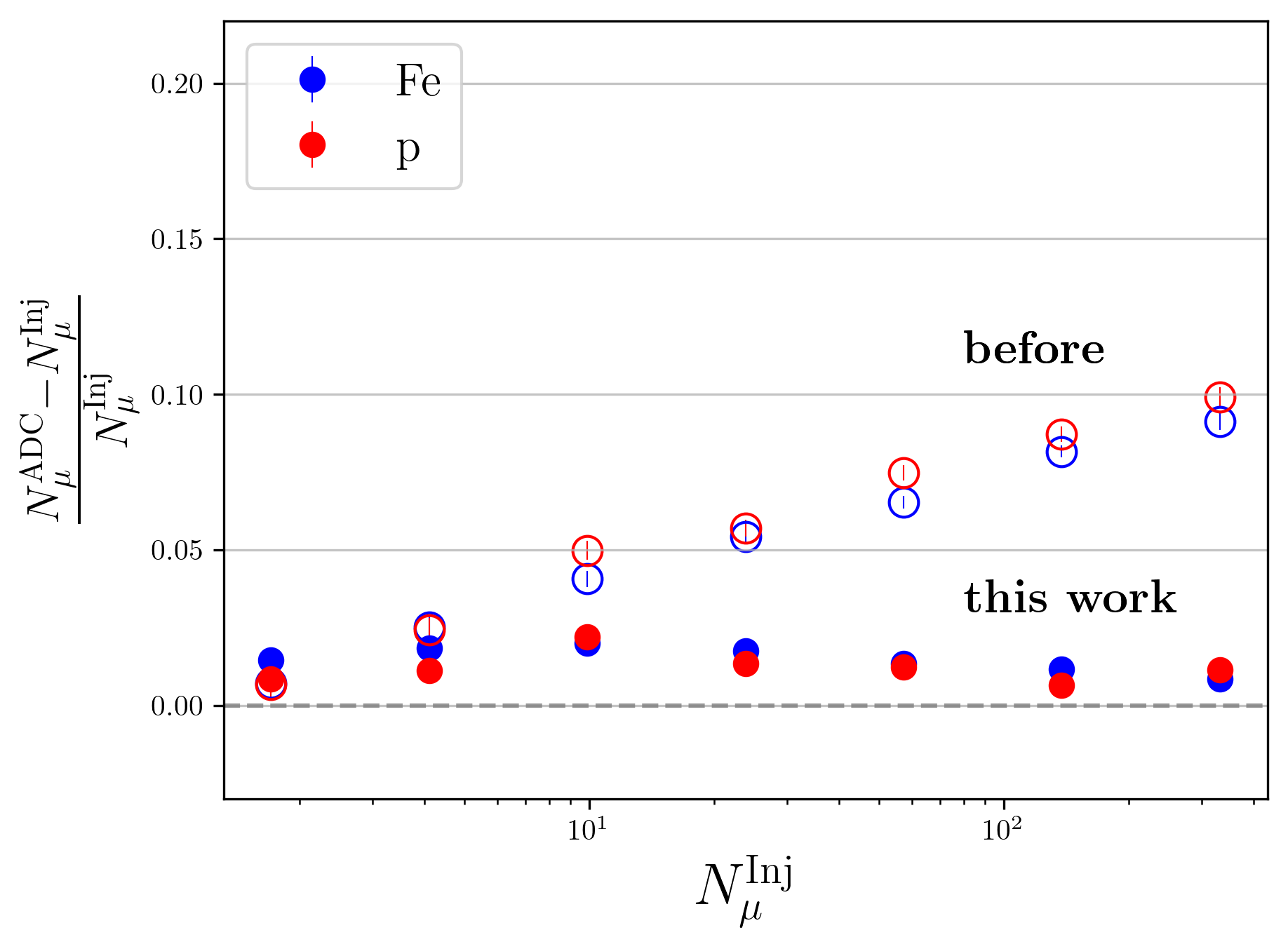}
        \caption{}
        \label{calibration_biasesa}
    \end{subfigure}

    \vspace{0.3cm}  % Space between rows

    \begin{subfigure}[b]{\w\textwidth}
        \includegraphics[width=\textwidth]{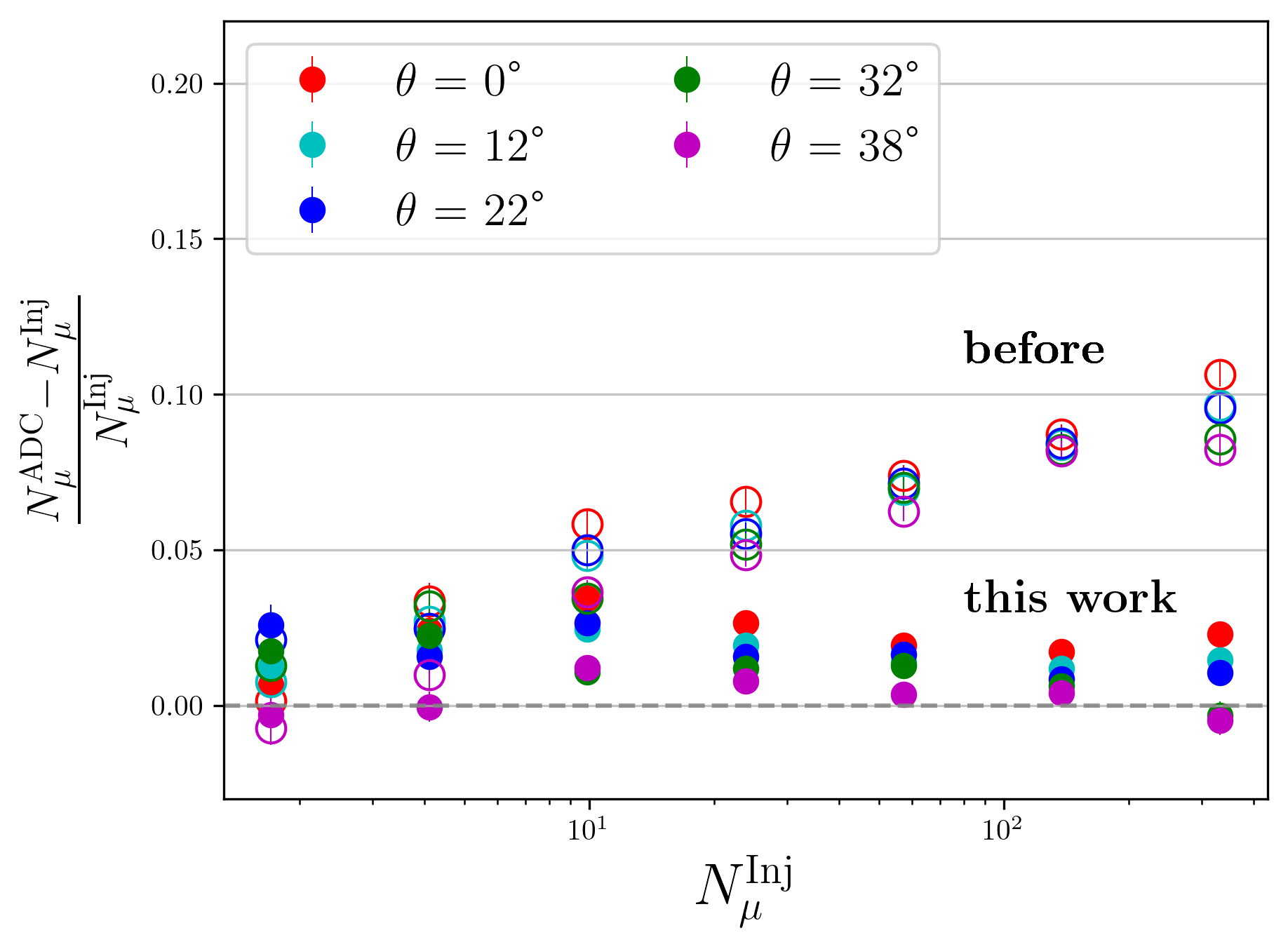}
        \caption{}
        \label{calibration_biasesc}
    \end{subfigure}
    \hfill
    \begin{subfigure}[b]{\w\textwidth}
        \includegraphics[width=\textwidth]{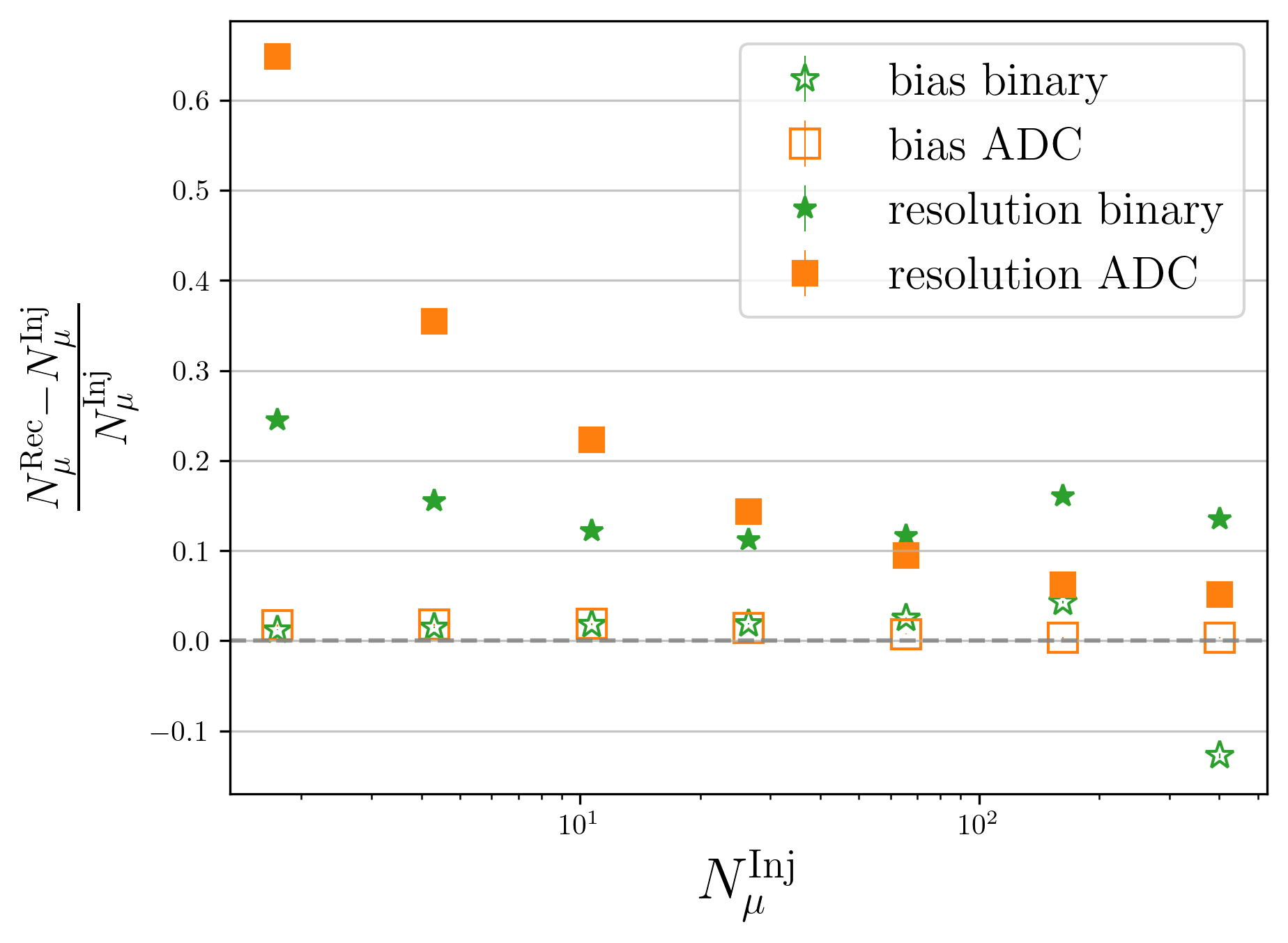}
        \caption{}
        \label{calibration_biasesd}
    \end{subfigure}
    \caption{Bias in muon reconstruction using the ADC mode for different (a) energies, (b) masses, and (c) zenith angles, comparing the calibration method (filled circles) with the previous approach (unfilled circles). (d) Bias and resolution in muon reconstruction for both the binary and ADC modes.}
  \label{calibration_biases}
\end{figure}

\begin{figure}[htbp]
\def\w{0.49}
    \centering
    \begin{subfigure}[b]{0.46\textwidth}
        \includegraphics[width=\textwidth]{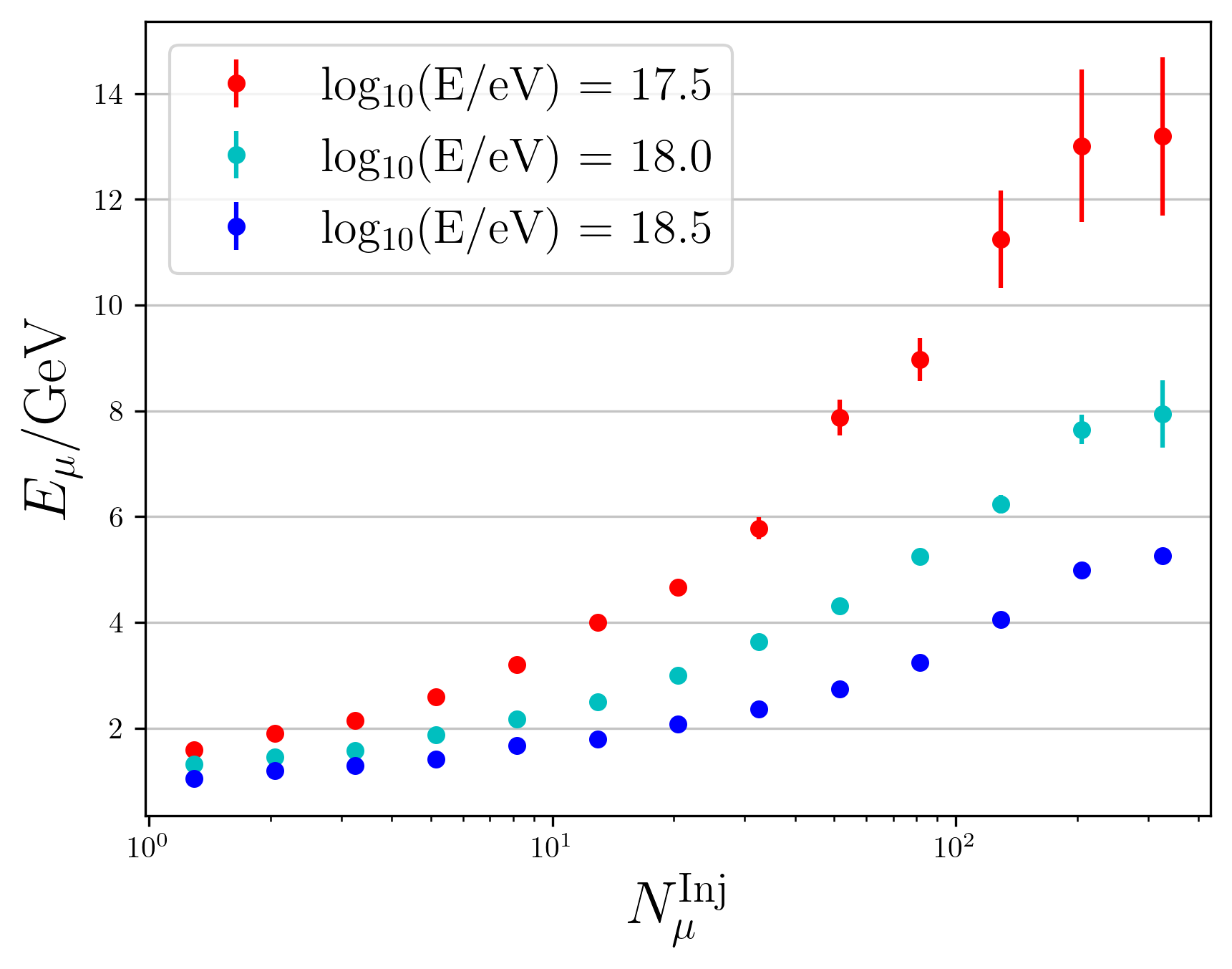}
        \caption{}
        \label{BiasADCEnergiasB}
    \end{subfigure}
    \hfill
    \begin{subfigure}[b]{\w\textwidth}
        \includegraphics[width=\textwidth]{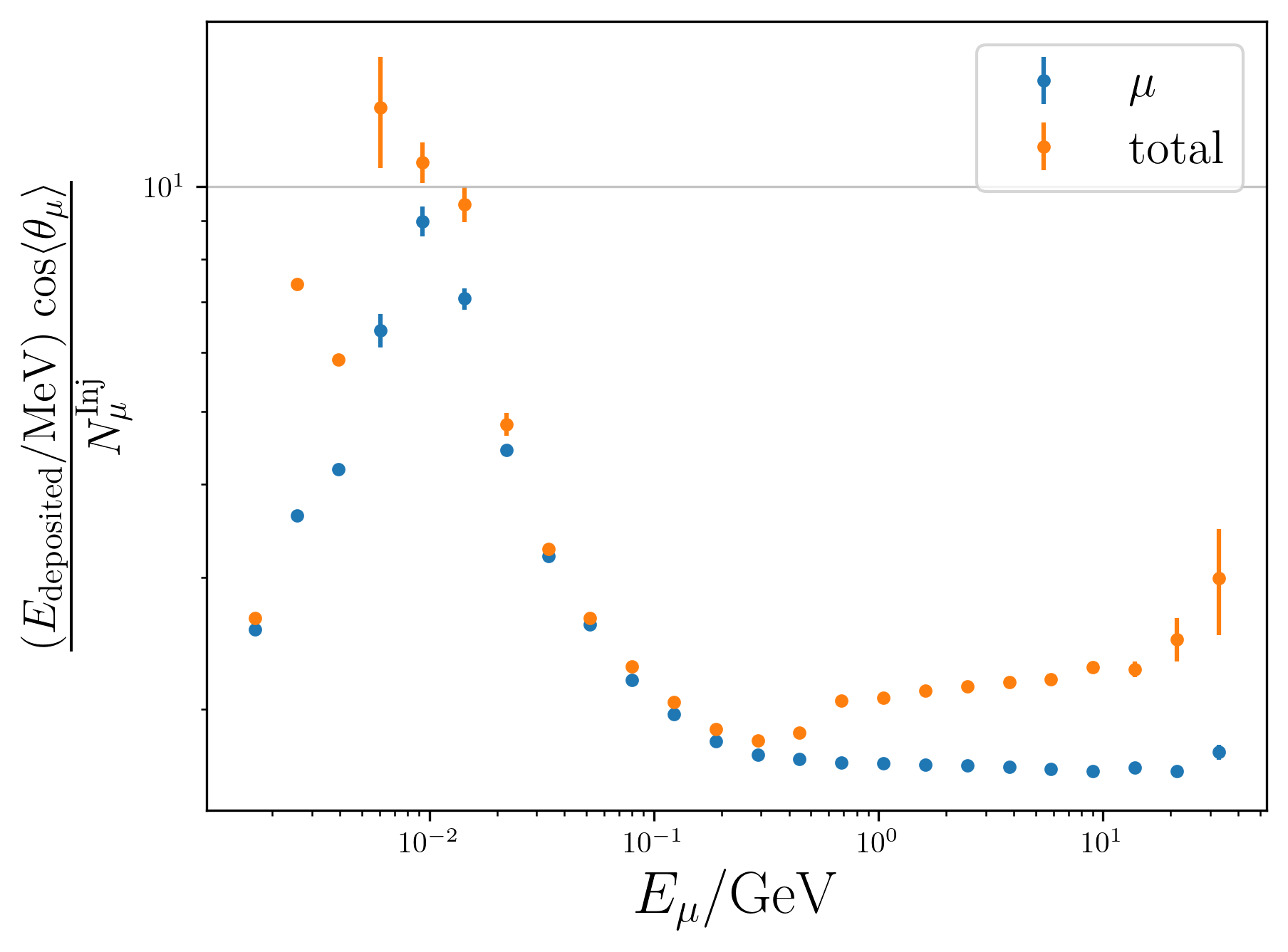}
        \caption{}
        \label{BiasADCEnergiasC}
    \end{subfigure}
    \caption{(a) Average muon energy as a function of injected muons per 10\,m$^2$ module. (b) Energy deposited per muon per vertical path length as a function of average muon energy, comparing muon-only and total contributions.}
  \label{BiasADCEnergias}
\end{figure}

The bias in the reconstructed muon number, computed using \cref{ecuacioncarga} under the assumption that the mean charge deposited by individual vertical muons is constant, is shown as unfilled circles in \cref{calibration_biasesb} for the different shower energies. The bias, displayed as a function of the number of injected muons per 10\,m$^2$ module ($N_\upmu^\text{Inj}$), increases with the muon number, reaching up to 20\% for showers of  $10^{17.5}$\,eV with $\sim$300 muons/10\,m$^2$. The bias is attributed to the increase in the energy deposition per muon with the muon energy, a dependence not captured by the constant calibration factor in \cref{ecuacioncarga}.  As shown in \cref{BiasADCEnergiasB}, the average muon energy, computed across all zenith angle bins, increases with the muon density and is higher for lower-energy showers for a given muon number. For muon energies above ${\sim}$0.3\,GeV, the total energy deposited per muon per vertical path length ($E^\text{total}_\text{deposited} \cos\avg{\theta_\upmu} \; / \; N_\upmu^\text{Inj}$) increases with the muon energy, while the purely muonic contribution remains nearly constant, as illustrated in \cref{BiasADCEnergiasC}. In the ADC mode, the measured charge reflects the total energy deposited in the scintillator, including contributions from secondary particles such as knock-on electrons, which are more likely to be produced and are more energetic as the kinetic energy of the muon increases. Within the range where the detector is not saturated, knock-on electrons dominate the non-muonic contributions, although protons and other particles like 
pions and kaons can also contribute slightly. For $10^{17.5}$\,eV showers at the highest muon density (${\sim}30$ muons/m$^2$), non-muonic particles contribute up to 31\% of the total deposited energy—23\% from knock-on electrons and 8\% from other particles~\cite{UHECRKnockOn}. Because these effects are more pronounced near the shower core, where higher-energy muons are concentrated, the resulting bias in $N_\upmu^\text{ADC}$ becomes strongly density-dependent, as seen in \cref{calibration_biasesb}. To address this bias, we propose in \cref{seccionimproved}  an improved calibration method based on a parametrisation that accounts for the energy dependence of the deposited charge.

\section{Improved calibration}\label{seccionimproved}

In this section, the same simulation procedure described in \cref{seccionknockon} was employed, with the addition of iron primaries to the dataset to study the composition dependence of the proposed calibration method. From these simulations, the charge deposited per muon per vertical path length, $q_{1\upmu}$, can be obtained for each 10\,m$^2$ UMD module by means of 
\begin{equation}
q_{1\upmu} = \frac{q_\text{meas}\cos\theta}{N_\upmu},
\label{ecuacq_1mu}
\end{equation}
using (1) $q_\text{meas}$ from the ADC mode, (2) $\cos(\theta)$ from the SD geometry reconstruction, and (3) $N_{\upmu}$. It is worth noting that \cref{ecuacq_1mu} can be seen as a reformulation of \cref{ecuacioncarga}, where the calibration factor—originally defined as the constant average charge deposited by simulating individual vertical muons—is now redefined as a parameterised quantity derived from shower events, incorporating information from multi-muon signals. The proposed calibration method involves estimating $q_{1\upmu}$ based on a measurable quantity ($m$) that correlates with the muon energy and is accessible even in high muon-density regions where the ADC mode is intended to be used. We select the distance to the shower axis, $r$, as the measurable quantity $m$, as it is directly related to the muon density. Two cases are considered in \cref{ecuacq_1mu}: the charge deposited per injected muon, $q_{1\upmu}^\text{Inj}$, computed using the true (Monte Carlo) number of injected muons to assess the validity of the method; and the reconstructed charge per muon, $q_{1\upmu}^\text{Rec}$, derived using the reconstructed muon number from the binary mode, $N_{\upmu}^\text{Bin}$, in the low-density regime, which can then be used to estimate higher muon densities by extrapolation.

\begin{figure}[htbp]
\def\w{0.49}
    \centering
    \begin{subfigure}[b]{\w\textwidth}
        \includegraphics[width=\textwidth]{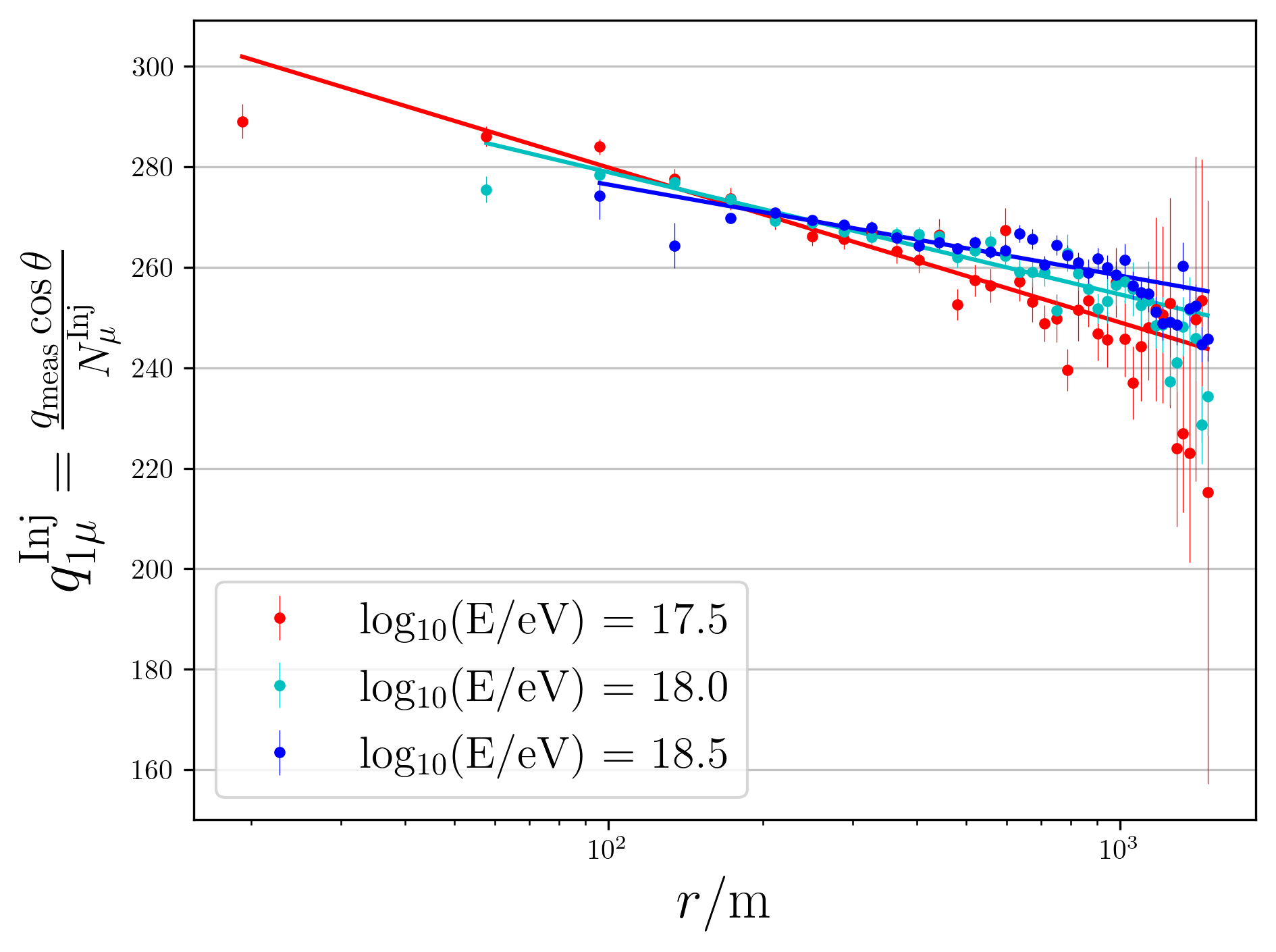}
        \caption{}
        \label{Dependences_of_charge_per_muonb}
    \end{subfigure}
    \hfill
    \begin{subfigure}[b]{\w\textwidth}
        \includegraphics[width=\textwidth]{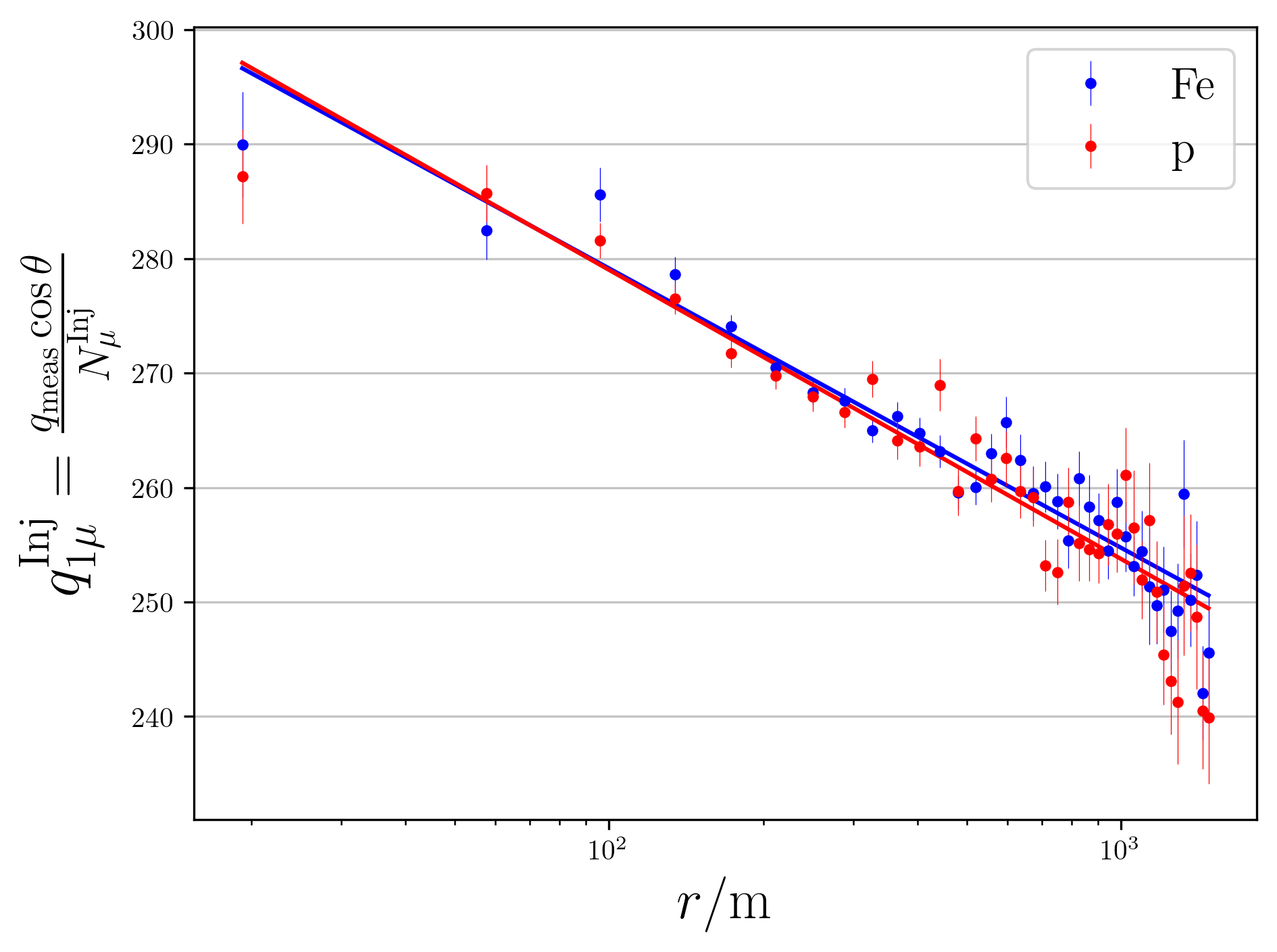}
        \caption{}
        \label{Dependences_of_charge_per_muona}
    \end{subfigure}

    \vspace{0.3cm}  % Space between rows

    \begin{subfigure}[b]{\w\textwidth}
        \includegraphics[width=\textwidth]{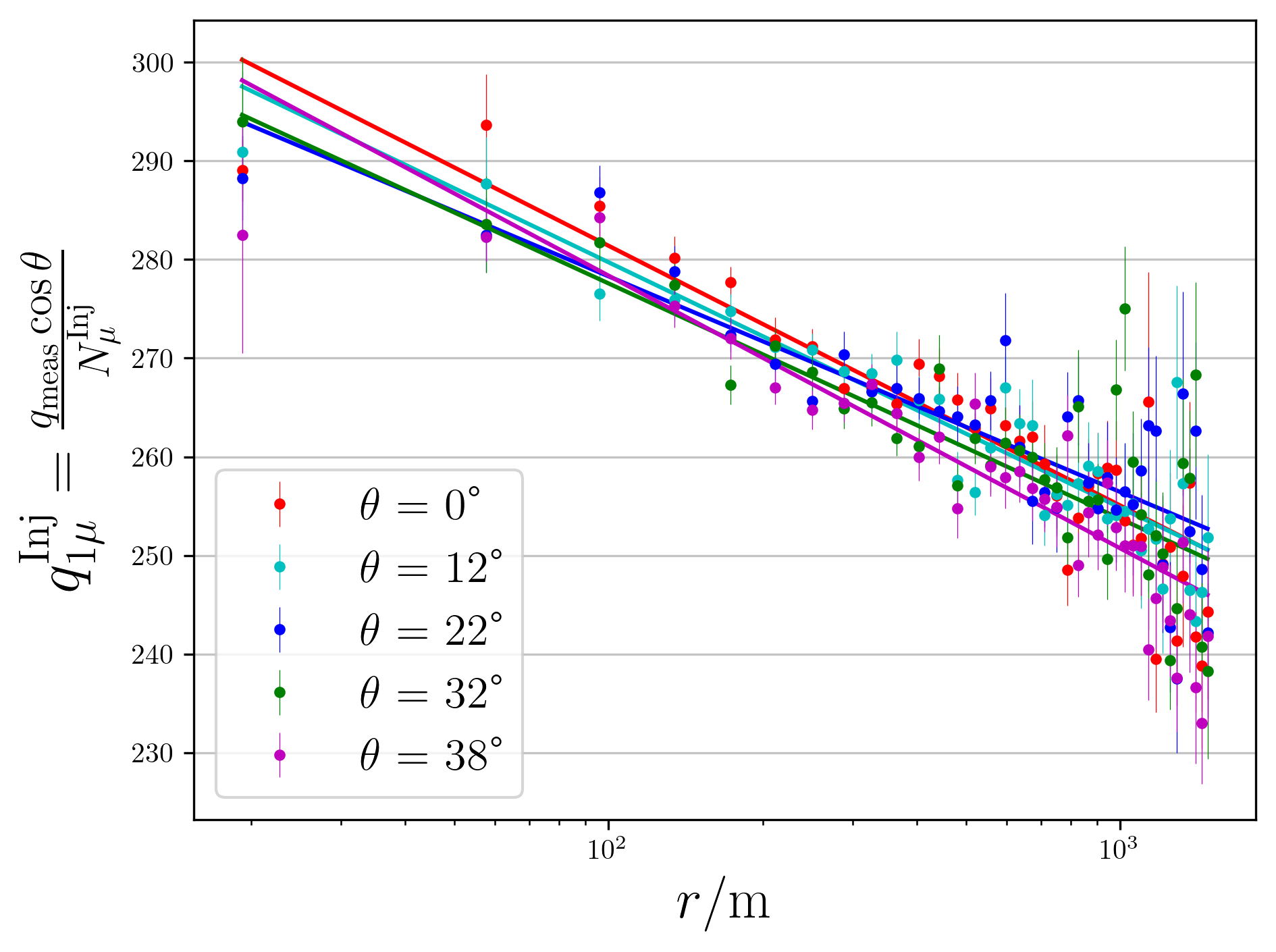}
        \caption{}
        \label{Dependences_of_charge_per_muonc}
    \end{subfigure}
    \hfill
    \begin{subfigure}[b]{\w\textwidth}
        \includegraphics[width=\textwidth]{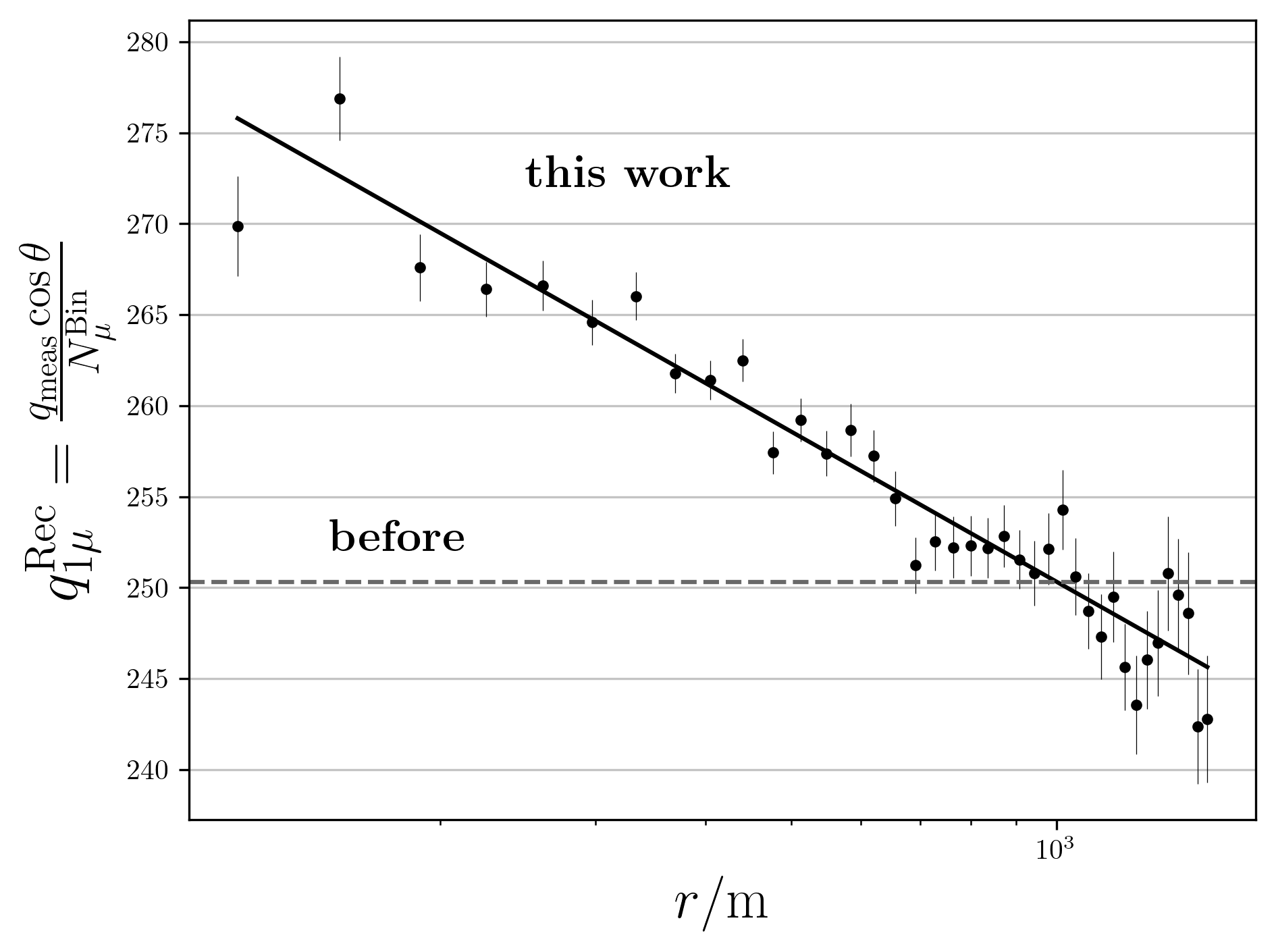}
        \caption{}
        \label{calibration_curve}
    \end{subfigure}
    \caption{
    Charge deposited per muon per vertical path length in a 10\,m$^2$ module as a function of the distance to the shower axis, discriminating by primary (a) energy, (b) mass, and (c) zenith angle. (d) Charge deposited per muon using the reconstructed number of muons in a 10\,m$^2$ module from the binary mode as a function of the distance to the shower axis.}
  \label{CalibrationDependencies}
\end{figure}

The charge deposited per muon, $q_{1\upmu}^\text{Inj}$, is shown as a function of the distance to the shower axis $r$  in \cref{Dependences_of_charge_per_muonb} for the three different energy bins, 
in \cref{Dependences_of_charge_per_muona} for the two different primaries,
and in \cref{Dependences_of_charge_per_muonc} for the five different zenith angle bins. A logarithmic dependence on $r$ is observed and parametrised as 
\begin{equation}
q_{1\upmu}(r)= A \log_{10} (r/\text{m}) + B,
\label{ecuacionparametrizacion2}
\end{equation}
with $A$ and $B$ fitting parameters. The solid lines represent the obtained fits, which performed very similarly, showing that the proposed method has no significant dependence on energy (less than 4\%), mass, and the incidence zenith angle (less than 2\%) of the primary.

The reconstructed charge per muon, $q_{1\upmu}^\text{Rec}$, derived from using in \cref{ecuacq_1mu} the reconstructed muon number $N_{\upmu}^\text{Bin}$ of the binary mode in the low-density regime ($<$70 muons/module), is shown in \cref{calibration_curve}, using a global fit to all data with cuts at $r{>}100$\,m. A cut of ${\sim}70$ muons per 10\,m$^2$ module was applied to ensure the binary mode operates within its optimal range (see \cref{calibration_biasesd}). Additionally, stations within $r {<} 100$\,m of the shower core were excluded to avoid inaccuracies in distance reconstruction. This global parametrisation is then used to reconstruct $N_{\upmu}^\text{ADC}$, by means of
\begin{equation}
N_\upmu^\text{ADC} (r) =
  \frac{q_\text{meas}\cos\theta}{q_{1\upmu}(r)}.
\label{ecuacioncargareformada}
\end{equation}

In \cref{calibration_biases}, the proposed method (filled circles) achieves a reconstruction bias below 5\% across all (a) energies, (b) masses, and (c) zenith angles, significantly improving upon the single-muon calibration method (unfilled circles) based on  \cref{ecuacioncarga}. The bias does not show a significant dependence on primary energy, mass, zenith angle, and muon number. In \cref{calibration_biasesd}, the bias and resolution in muon reconstruction for both the binary and ADC modes are shown. The results were obtained using the full dataset. The binary mode provides the best performance at low muon densities, while the ADC mode is more accurate at intermediate and high densities. For muon counts above ${\sim}70$ muons/module, the ADC mode yields superior bias and resolution.

The calibration method was applied to the UMD data in a six-year period. The data utilized were divided into two different periods depending on the electronics of the SD array: Phase I, i.e.\ acquired using the former electronics from 2019 to 2022, and Phase II, i.e.\ acquired using the new electronics, between 2023 and September of 2024. Calibration curves were obtained for each period, covering 76 modules of 10\,m$^2$ in Phase I and 99 modules of 10\,m$^2$ in Phase II. Each module was calibrated individually. As is customary for the UMD, the showers reconstructed were required to have energies larger than  $10^{17.3}$\,eV and zenith angles smaller than 45$^\circ$. 

\begin{figure}[htbp]
\def\w{0.49}
    \centering
        \begin{subfigure}[b]{\w\textwidth}
        \includegraphics[width=\textwidth]{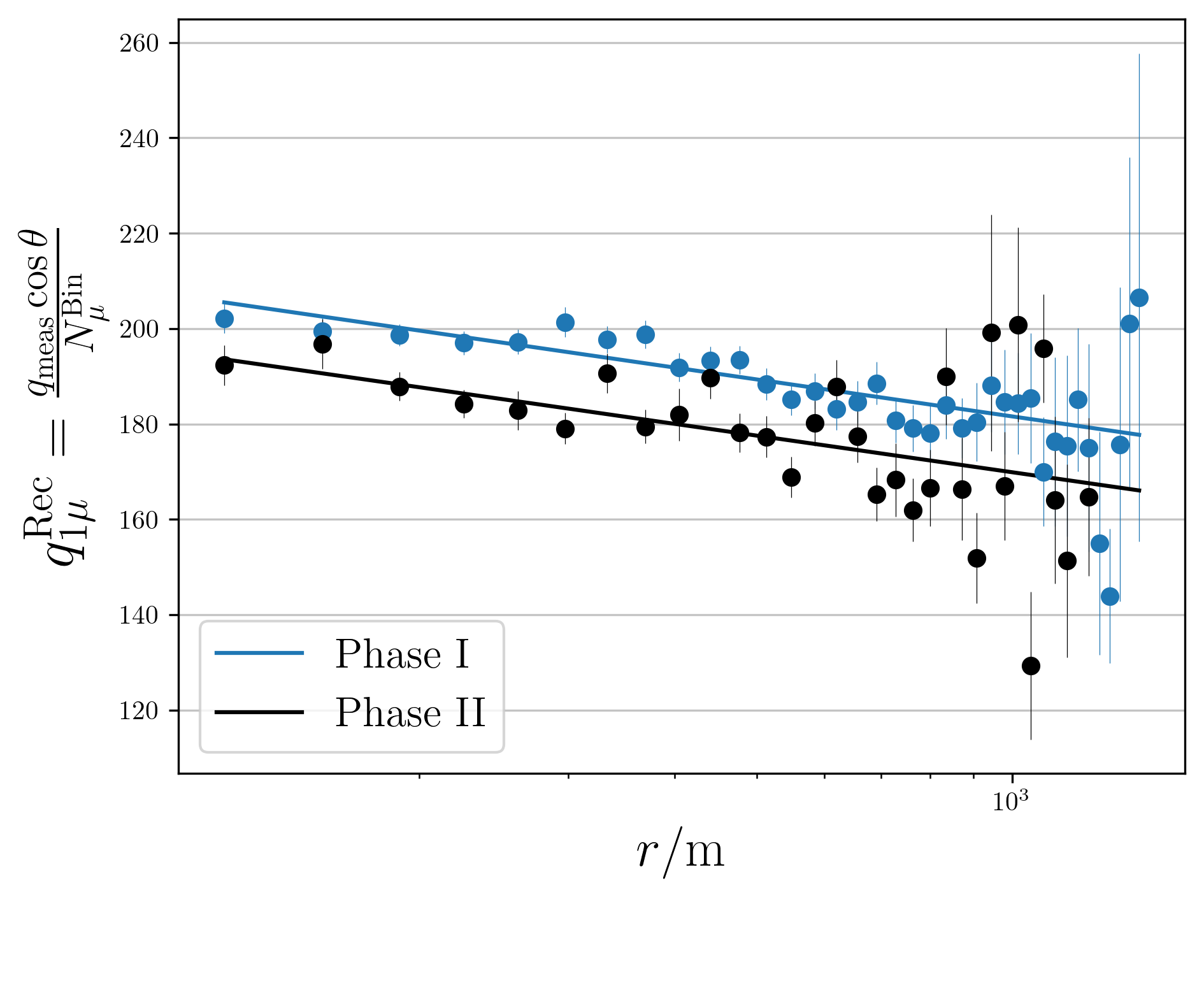}
        \caption{}
        \label{calibration_datosa}
    \end{subfigure}
    \hfill
    \begin{subfigure}[b]{\w\textwidth}
        \includegraphics[width=\textwidth]{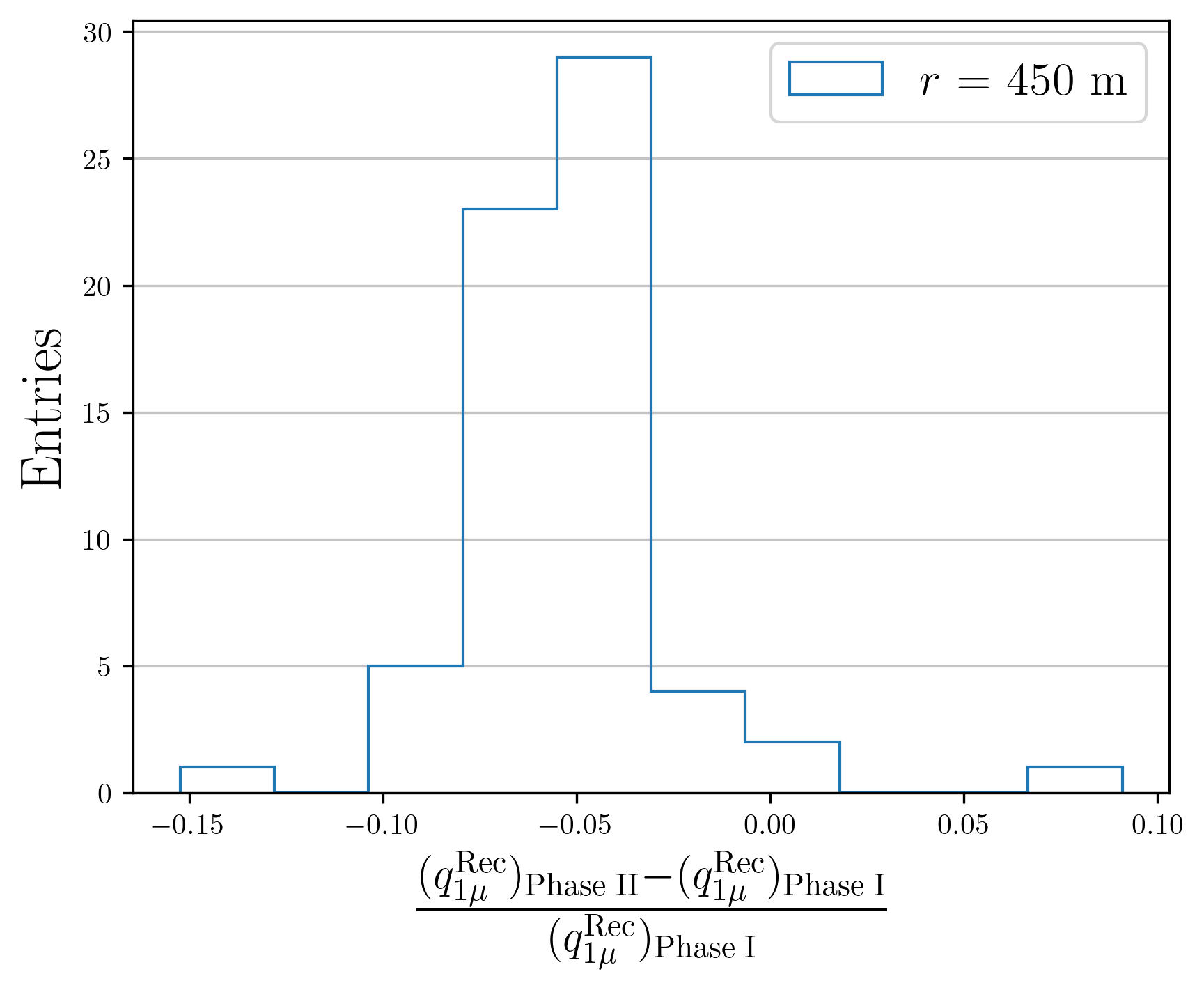}
        \caption{}
        \label{calibration_datosb}
    \end{subfigure}
    \caption{(a) An example of the calibration method obtained for Phase II (black) and Phase I (blue) for an individual UMD station. (b) Relative difference of the calibration curve evaluated at $r$=450\,m between Phase II and Phase I.}
  \label{calibration_datos}
\end{figure}

In  \cref{calibration_datosa}, an example of the calibration method is shown for an individual UMD station, illustrating the calibration curves for Phase I (blue) and Phase II (black). The charge deposited per muon increases at small core-distances due to the energy deposition of the knock-on electrons, reproducing the effect observed in simulations. The average values of the parametrisation coefficient $A$, obtained by averaging over all deployed modules, were $A = (-28.8 \pm 0.9)$ a.u.\ (arbitrary units) in Phase I and $A = (-26.1 \pm 0.9)$ a.u.\ in Phase II, compared to $A = (-27.4 \pm 1.1)$ a.u.\ in simulations. The close agreement indicates that simulations reproduce the increase in the charge deposited per muon with the particle density, as described by \cref{ecuacionparametrizacion2}, and accurately account for the contribution of knock-on electrons.

Due to the extended deployment timeline and uneven data availability across modules, the calibration stability could only be evaluated in a subset of stations with sufficient statistics during Phase I. In these cases, a long-term performance effect was observed, with lower measured charges in the later period. An ageing rate of $\sim$-2.5\% per year for the ADC mode was previously reported in Ref.~\cite{joaquinICRC}. This trend is consistent with the ageing observed in similar scintillator detectors, such as those in the MINOS experiment~\cite{minos}, and should be considered as a systematic uncertainty in Phase I analyses. This long-term performance effect was also confirmed with higher statistics in Phase II, as observed in \Cref{calibration_datosa}. \Cref{calibration_datosb} shows the relative difference in the average charge measured at the optimal reference distance for composition studies ($r = 450$\,m), with Phase II values being approximately 5\% lower.

\section{Summary}

The calibration of underground muon detectors based on plastic scintillators using the average charge from single vertical muons can lead to biases in the reconstructed muon density.  For the ADC mode of the UMD, this bias can reach up to 20\% in showers of $10^{17.5}$\,eV at densities of ${\sim}30\,\upmu/\text{m}^2$. The bias is attributed to the increased energy deposition per muon, especially near the shower axis, where high-energy muons produce more secondary knock-on electrons that are generated in the surrounding soil. These knock-on electrons dominate the non-muonic contribution to the signal and account for up to 23\% of the total energy deposition in extreme cases. To address this, a new calibration method was developed, based on parametrising the charge per muon per vertical path length using the binary mode of the UMD and the distance to the shower core for low muon densities, and extrapolating it to high muon densities. The method significantly reduces the reconstruction bias to less than 5\% and shows no significant dependence on energy, zenith angle, or primary mass.

The method was applied to both Phase I (2019–2022) and Phase II (2023–2024) data, yielding calibration curves for 76 and 99 modules of 10\,m$^2$, respectively. The calibration parameter $A$, which relates the charge increase with muon density, showed good agreement between data and simulations, confirming that the impact of knock-on electrons is well-modelled. Additionally, a long-term performance effect was observed, with Phase II measurements showing a $\sim$5\% reduction in the average charge at $r=450$\,m compared to Phase I, consistent with the ageing effects reported in similar scintillator detectors.

\newpage

\section*{The Pierre Auger Collaboration}

{\footnotesize\setlength{\baselineskip}{10pt}
\noindent
\begin{wrapfigure}[11]{l}{0.12\linewidth}
\vspace{-4pt}
\includegraphics[width=0.95\linewidth]{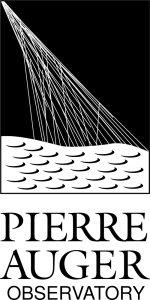}
\end{wrapfigure}
\begin{sloppypar}\noindent
% created on 2025-06-06
A.~Abdul Halim$^{13}$,
P.~Abreu$^{70}$,
M.~Aglietta$^{53,51}$,
I.~Allekotte$^{1}$,
K.~Almeida Cheminant$^{78,77}$,
A.~Almela$^{7,12}$,
R.~Aloisio$^{44,45}$,
J.~Alvarez-Mu\~niz$^{76}$,
A.~Ambrosone$^{44}$,
J.~Ammerman Yebra$^{76}$
% ...

\end{sloppypar}
\begin{center}
\end{center}

\vspace{1ex}
% created on 2025-06-06
% needs \usepackage{enumitem}
\begin{description}[labelsep=0.2em,align=right,labelwidth=0.7em,labelindent=0em,leftmargin=2em,noitemsep,before={\renewcommand\makelabel[1]{##1 }}]
\item[$^{1}$] Centro At\'omico Bariloche and Instituto Balseiro (CNEA-UNCuyo-CONICET), San Carlos de Bariloche, Argentina
\item[$^{2}$] Departamento de F\'\i{}sica and Departamento de Ciencias de la Atm\'osfera y los Oc\'eanos, FCEyN, Universidad de Buenos Aires and CONICET, Buenos Aires, Argentina
\item[$^{3}$] IFLP, Universidad Nacional de La Plata and CONICET, La Plata, Argentina
\item[$^{4}$] Instituto de Astronom\'\i{}a y F\'\i{}sica del Espacio (IAFE, CONICET-UBA), Buenos Aires, Argentina
\item[$^{5}$] Instituto de F\'\i{}sica de Rosario (IFIR) -- CONICET/U.N.R.\ and Facultad de Ciencias Bioqu\'\i{}micas y Farmac\'euticas U.N.R., Rosario, Argentina
\item[$^{6}$] Instituto de Tecnolog\'\i{}as en Detecci\'on y Astropart\'\i{}culas (CNEA, CONICET, UNSAM), and Universidad Tecnol\'ogica Nacional -- Facultad Regional Mendoza (CONICET/CNEA), Mendoza, Argentina
\item[$^{7}$] Instituto de Tecnolog\'\i{}as en Detecci\'on y Astropart\'\i{}culas (CNEA, CONICET, UNSAM), Buenos Aires, Argentina
\item[$^{8}$] International Center of Advanced Studies and Instituto de Ciencias F\'\i{}sicas, ECyT-UNSAM and CONICET, Campus Miguelete -- San Mart\'\i{}n, Buenos Aires, Argentina
\item[$^{9}$] Laboratorio Atm\'osfera -- Departamento de Investigaciones en L\'aseres y sus Aplicaciones -- UNIDEF (CITEDEF-CONICET), Argentina
\item[$^{10}$] Observatorio Pierre Auger, Malarg\"ue, Argentina
\item[$^{11}$] Observatorio Pierre Auger and Comisi\'on Nacional de Energ\'\i{}a At\'omica, Malarg\"ue, Argentina
\item[$^{12}$] Universidad Tecnol\'ogica Nacional -- Facultad Regional Buenos Aires, Buenos Aires, Argentina
\item[$^{13}$] University of Adelaide, Adelaide, S.A., Australia
\item[$^{14}$] Universit\'e Libre de Bruxelles (ULB), Brussels, Belgium
\item[$^{15}$] Vrije Universiteit Brussels, Brussels, Belgium
\item[$^{16}$] Centro Brasileiro de Pesquisas Fisicas, Rio de Janeiro, RJ, Brazil
\item[$^{17}$] Centro Federal de Educa\c{c}\~ao Tecnol\'ogica Celso Suckow da Fonseca, Petropolis, Brazil
\item[$^{18}$] Instituto Federal de Educa\c{c}\~ao, Ci\^encia e Tecnologia do Rio de Janeiro (IFRJ), Brazil
\item[$^{19}$] Universidade de S\~ao Paulo, Escola de Engenharia de Lorena, Lorena, SP, Brazil
\item[$^{20}$] Universidade de S\~ao Paulo, Instituto de F\'\i{}sica de S\~ao Carlos, S\~ao Carlos, SP, Brazil
\item[$^{21}$] Universidade de S\~ao Paulo, Instituto de F\'\i{}sica, S\~ao Paulo, SP, Brazil
\item[$^{22}$] Universidade Estadual de Campinas (UNICAMP), IFGW, Campinas, SP, Brazil
\item[$^{23}$] Universidade Estadual de Feira de Santana, Feira de Santana, Brazil
\item[$^{24}$] Universidade Federal de Campina Grande, Centro de Ciencias e Tecnologia, Campina Grande, Brazil
\item[$^{25}$] Universidade Federal do ABC, Santo Andr\'e, SP, Brazil
\item[$^{26}$] Universidade Federal do Paran\'a, Setor Palotina, Palotina, Brazil
\item[$^{27}$] Universidade Federal do Rio de Janeiro, Instituto de F\'\i{}sica, Rio de Janeiro, RJ, Brazil
\item[$^{28}$] Universidad de Medell\'\i{}n, Medell\'\i{}n, Colombia
\item[$^{29}$] Universidad Industrial de Santander, Bucaramanga, Colombia
\item[$^{30}$] Charles University, Faculty of Mathematics and Physics, Institute of Particle and Nuclear Physics, Prague, Czech Republic
\item[$^{31}$] Institute of Physics of the Czech Academy of Sciences, Prague, Czech Republic
\item[$^{32}$] Palacky University, Olomouc, Czech Republic
\item[$^{33}$] CNRS/IN2P3, IJCLab, Universit\'e Paris-Saclay, Orsay, France
\item[$^{34}$] Laboratoire de Physique Nucl\'eaire et de Hautes Energies (LPNHE), Sorbonne Universit\'e, Universit\'e de Paris, CNRS-IN2P3, Paris, France
\item[$^{35}$] Univ.\ Grenoble Alpes, CNRS, Grenoble Institute of Engineering Univ.\ Grenoble Alpes, LPSC-IN2P3, 38000 Grenoble, France
\item[$^{36}$] Universit\'e Paris-Saclay, CNRS/IN2P3, IJCLab, Orsay, France
\item[$^{37}$] Bergische Universit\"at Wuppertal, Department of Physics, Wuppertal, Germany
\item[$^{38}$] Karlsruhe Institute of Technology (KIT), Institute for Experimental Particle Physics, Karlsruhe, Germany
\item[$^{39}$] Karlsruhe Institute of Technology (KIT), Institut f\"ur Prozessdatenverarbeitung und Elektronik, Karlsruhe, Germany
\item[$^{40}$] Karlsruhe Institute of Technology (KIT), Institute for Astroparticle Physics, Karlsruhe, Germany
\item[$^{41}$] RWTH Aachen University, III.\ Physikalisches Institut A, Aachen, Germany
\item[$^{42}$] Universit\"at Hamburg, II.\ Institut f\"ur Theoretische Physik, Hamburg, Germany
\item[$^{43}$] Universit\"at Siegen, Department Physik -- Experimentelle Teilchenphysik, Siegen, Germany
\item[$^{44}$] Gran Sasso Science Institute, L'Aquila, Italy
\item[$^{45}$] INFN Laboratori Nazionali del Gran Sasso, Assergi (L'Aquila), Italy
\item[$^{46}$] INFN, Sezione di Catania, Catania, Italy
\item[$^{47}$] INFN, Sezione di Lecce, Lecce, Italy
\item[$^{48}$] INFN, Sezione di Milano, Milano, Italy
\item[$^{49}$] INFN, Sezione di Napoli, Napoli, Italy
\item[$^{50}$] INFN, Sezione di Roma ``Tor Vergata'', Roma, Italy
\item[$^{51}$] INFN, Sezione di Torino, Torino, Italy
\item[$^{52}$] Istituto di Astrofisica Spaziale e Fisica Cosmica di Palermo (INAF), Palermo, Italy
\item[$^{53}$] Osservatorio Astrofisico di Torino (INAF), Torino, Italy
\item[$^{54}$] Politecnico di Milano, Dipartimento di Scienze e Tecnologie Aerospaziali , Milano, Italy
\item[$^{55}$] Universit\`a del Salento, Dipartimento di Matematica e Fisica ``E.\ De Giorgi'', Lecce, Italy
\item[$^{56}$] Universit\`a dell'Aquila, Dipartimento di Scienze Fisiche e Chimiche, L'Aquila, Italy
\item[$^{57}$] Universit\`a di Catania, Dipartimento di Fisica e Astronomia ``Ettore Majorana``, Catania, Italy
\item[$^{58}$] Universit\`a di Milano, Dipartimento di Fisica, Milano, Italy
\item[$^{59}$] Universit\`a di Napoli ``Federico II'', Dipartimento di Fisica ``Ettore Pancini'', Napoli, Italy
\item[$^{60}$] Universit\`a di Palermo, Dipartimento di Fisica e Chimica ''E.\ Segr\`e'', Palermo, Italy
\item[$^{61}$] Universit\`a di Roma ``Tor Vergata'', Dipartimento di Fisica, Roma, Italy
\item[$^{62}$] Universit\`a Torino, Dipartimento di Fisica, Torino, Italy
\item[$^{63}$] Benem\'erita Universidad Aut\'onoma de Puebla, Puebla, M\'exico
\item[$^{64}$] Unidad Profesional Interdisciplinaria en Ingenier\'\i{}a y Tecnolog\'\i{}as Avanzadas del Instituto Polit\'ecnico Nacional (UPIITA-IPN), M\'exico, D.F., M\'exico
\item[$^{65}$] Universidad Aut\'onoma de Chiapas, Tuxtla Guti\'errez, Chiapas, M\'exico
\item[$^{66}$] Universidad Michoacana de San Nicol\'as de Hidalgo, Morelia, Michoac\'an, M\'exico
\item[$^{67}$] Universidad Nacional Aut\'onoma de M\'exico, M\'exico, D.F., M\'exico
\item[$^{68}$] Institute of Nuclear Physics PAN, Krakow, Poland
\item[$^{69}$] University of \L{}\'od\'z, Faculty of High-Energy Astrophysics,\L{}\'od\'z, Poland
\item[$^{70}$] Laborat\'orio de Instrumenta\c{c}\~ao e F\'\i{}sica Experimental de Part\'\i{}culas -- LIP and Instituto Superior T\'ecnico -- IST, Universidade de Lisboa -- UL, Lisboa, Portugal
\item[$^{71}$] ``Horia Hulubei'' National Institute for Physics and Nuclear Engineering, Bucharest-Magurele, Romania
\item[$^{72}$] Institute of Space Science, Bucharest-Magurele, Romania
\item[$^{73}$] Center for Astrophysics and Cosmology (CAC), University of Nova Gorica, Nova Gorica, Slovenia
\item[$^{74}$] Experimental Particle Physics Department, J.\ Stefan Institute, Ljubljana, Slovenia
\item[$^{75}$] Universidad de Granada and C.A.F.P.E., Granada, Spain
\item[$^{76}$] Instituto Galego de F\'\i{}sica de Altas Enerx\'\i{}as (IGFAE), Universidade de Santiago de Compostela, Santiago de Compostela, Spain
\item[$^{77}$] IMAPP, Radboud University Nijmegen, Nijmegen, The Netherlands
\item[$^{78}$] Nationaal Instituut voor Kernfysica en Hoge Energie Fysica (NIKHEF), Science Park, Amsterdam, The Netherlands
\item[$^{79}$] Stichting Astronomisch Onderzoek in Nederland (ASTRON), Dwingeloo, The Netherlands
\item[$^{80}$] Universiteit van Amsterdam, Faculty of Science, Amsterdam, The Netherlands
\item[$^{81}$] Case Western Reserve University, Cleveland, OH, USA
\item[$^{82}$] Colorado School of Mines, Golden, CO, USA
\item[$^{83}$] Department of Physics and Astronomy, Lehman College, City University of New York, Bronx, NY, USA
\item[$^{84}$] Michigan Technological University, Houghton, MI, USA
\item[$^{85}$] New York University, New York, NY, USA
\item[$^{86}$] University of Chicago, Enrico Fermi Institute, Chicago, IL, USA
\item[$^{87}$] University of Delaware, Department of Physics and Astronomy, Bartol Research Institute, Newark, DE, USA
\item[] -----
\item[$^{a}$] Max-Planck-Institut f\"ur Radioastronomie, Bonn, Germany
\item[$^{b}$] also at Kapteyn Institute, University of Groningen, Groningen, The Netherlands
\item[$^{c}$] School of Physics and Astronomy, University of Leeds, Leeds, United Kingdom
\item[$^{d}$] Fermi National Accelerator Laboratory, Fermilab, Batavia, IL, USA
\item[$^{e}$] Pennsylvania State University, University Park, PA, USA
\item[$^{f}$] Colorado State University, Fort Collins, CO, USA
\item[$^{g}$] Louisiana State University, Baton Rouge, LA, USA
\item[$^{h}$] now at Graduate School of Science, Osaka Metropolitan University, Osaka, Japan
\item[$^{i}$] Institut universitaire de France (IUF), France
\item[$^{j}$] now at Technische Universit\"at Dortmund and Ruhr-Universit\"at Bochum, Dortmund and Bochum, Germany
\end{description}

% created on 2025-06-06
\section*{Acknowledgments}

\begin{sloppypar}
The successful installation, commissioning, and operation of the Pierre
Auger Observatory would not have been possible without the strong
commitment and effort from the technical and administrative staff in
Malarg\"ue. We are very grateful to the following agencies and
organizations for financial support:
\end{sloppypar}

\begin{sloppypar}
Argentina -- Comisi\'on Nacional de Energ\'\i{}a At\'omica; Agencia Nacional de
Promoci\'on Cient\'\i{}fica y Tecnol\'ogica (ANPCyT); Consejo Nacional de
Investigaciones Cient\'\i{}ficas y T\'ecnicas (CONICET); Gobierno de la
Provincia de Mendoza; Municipalidad de Malarg\"ue; NDM Holdings and Valle
Las Le\~nas; in gratitude for their continuing cooperation over land
access; Australia -- the Australian Research Council; Belgium -- Fonds
de la Recherche Scientifique (FNRS); Research Foundation Flanders (FWO),
Marie Curie Action of the European Union Grant No.~101107047; Brazil --
Conselho Nacional de Desenvolvimento Cient\'\i{}fico e Tecnol\'ogico (CNPq);
Financiadora de Estudos e Projetos (FINEP); Funda\c{c}\~ao de Amparo \`a
Pesquisa do Estado de Rio de Janeiro (FAPERJ); S\~ao Paulo Research
Foundation (FAPESP) Grants No.~2019/10151-2, No.~2010/07359-6 and
No.~1999/05404-3; Minist\'erio da Ci\^encia, Tecnologia, Inova\c{c}\~oes e
Comunica\c{c}\~oes (MCTIC); Czech Republic -- GACR 24-13049S, CAS LQ100102401,
MEYS LM2023032, CZ.02.1.01/0.0/0.0/16{\textunderscore}013/0001402,
CZ.02.1.01/0.0/0.0/18{\textunderscore}046/0016010 and
CZ.02.1.01/0.0/0.0/17{\textunderscore}049/0008422 and CZ.02.01.01/00/22{\textunderscore}008/0004632;
France -- Centre de Calcul IN2P3/CNRS; Centre National de la Recherche
Scientifique (CNRS); Conseil R\'egional Ile-de-France; D\'epartement
Physique Nucl\'eaire et Corpusculaire (PNC-IN2P3/CNRS); D\'epartement
Sciences de l'Univers (SDU-INSU/CNRS); Institut Lagrange de Paris (ILP)
Grant No.~LABEX ANR-10-LABX-63 within the Investissements d'Avenir
Programme Grant No.~ANR-11-IDEX-0004-02; Germany -- Bundesministerium
f\"ur Bildung und Forschung (BMBF); Deutsche Forschungsgemeinschaft (DFG);
Finanzministerium Baden-W\"urttemberg; Helmholtz Alliance for
Astroparticle Physics (HAP); Helmholtz-Gemeinschaft Deutscher
Forschungszentren (HGF); Ministerium f\"ur Kultur und Wissenschaft des
Landes Nordrhein-Westfalen; Ministerium f\"ur Wissenschaft, Forschung und
Kunst des Landes Baden-W\"urttemberg; Italy -- Istituto Nazionale di
Fisica Nucleare (INFN); Istituto Nazionale di Astrofisica (INAF);
Ministero dell'Universit\`a e della Ricerca (MUR); CETEMPS Center of
Excellence; Ministero degli Affari Esteri (MAE), ICSC Centro Nazionale
di Ricerca in High Performance Computing, Big Data and Quantum
Computing, funded by European Union NextGenerationEU, reference code
CN{\textunderscore}00000013; M\'exico -- Consejo Nacional de Ciencia y Tecnolog\'\i{}a
(CONACYT) No.~167733; Universidad Nacional Aut\'onoma de M\'exico (UNAM);
PAPIIT DGAPA-UNAM; The Netherlands -- Ministry of Education, Culture and
Science; Netherlands Organisation for Scientific Research (NWO); Dutch
national e-infrastructure with the support of SURF Cooperative; Poland
-- Ministry of Education and Science, grants No.~DIR/WK/2018/11 and
2022/WK/12; National Science Centre, grants No.~2016/22/M/ST9/00198,
2016/23/B/ST9/01635, 2020/39/B/ST9/01398, and 2022/45/B/ST9/02163;
Portugal -- Portuguese national funds and FEDER funds within Programa
Operacional Factores de Competitividade through Funda\c{c}\~ao para a Ci\^encia
e a Tecnologia (COMPETE); Romania -- Ministry of Research, Innovation
and Digitization, CNCS-UEFISCDI, contract no.~30N/2023 under Romanian
National Core Program LAPLAS VII, grant no.~PN 23 21 01 02 and project
number PN-III-P1-1.1-TE-2021-0924/TE57/2022, within PNCDI III; Slovenia
-- Slovenian Research Agency, grants P1-0031, P1-0385, I0-0033, N1-0111;
Spain -- Ministerio de Ciencia e Innovaci\'on/Agencia Estatal de
Investigaci\'on (PID2019-105544GB-I00, PID2022-140510NB-I00 and
RYC2019-027017-I), Xunta de Galicia (CIGUS Network of Research Centers,
Consolidaci\'on 2021 GRC GI-2033, ED431C-2021/22 and ED431F-2022/15),
Junta de Andaluc\'\i{}a (SOMM17/6104/UGR and P18-FR-4314), and the European
Union (Marie Sklodowska-Curie 101065027 and ERDF); USA -- Department of
Energy, Contracts No.~DE-AC02-07CH11359, No.~DE-FR02-04ER41300,
No.~DE-FG02-99ER41107 and No.~DE-SC0011689; National Science Foundation,
Grant No.~0450696, and NSF-2013199; The Grainger Foundation; Marie
Curie-IRSES/EPLANET; European Particle Physics Latin American Network;
and UNESCO.
\end{sloppypar}

}


\begin{thebibliography}{99}

\footnotesize
\raggedright
\setlength{\itemsep}{0pt}
\def\vyp#1#2#3{\textbf{#1} (#2) #3} % vyp = volume year page

\bibitem{upgrade}
The Pierre Auger Collaboration, Science Reviews \vyp{1}{2020}{8}.

\bibitem{umddesign1}
The Pierre Auger Collaboration, JINST \vyp{16}{2021}{P01026.}

\bibitem{flavia}
F.~Gesualdi and D.~Supanitsky, Eur.\ Phys.\ J.\ C, \vyp{82}{2022}{925}.

\bibitem{joaquincorner}
J.~de Jes\'us, J.~M.~Figueira, F.~Sánchez, and D.~Veberi\v{c}, PoS \vyp{UHECR2024}{2024}{078.}

\bibitem{calibrationADCpaper}
The Pierre Auger Collaboration, JINST \vyp{16}{2021}{P04003.}

\bibitem{knockonelectrons}
C.~Anderson, G.~McKinney, J.~Tutt, and M.~James, Phys.\ Proc.\ \vyp{90}{2017}{229-236.}

\bibitem{UHECRKnockOn}
M.~Scornavacche, J.~M.~Figueira, F.~Sánchez, and D.~Veberi\v{c}, PoS \vyp{UHECR2024}{2024}{116.}

\bibitem{Geant4}
S.~Agostinelli \emph{et al.}, Nucl.\ Instrum.\ Meth.\ A \vyp{506}{2003}{250--303.}

\bibitem{offline}
S.~Argiro  \emph{et al.}, Nucl.\ Instrum.\ Meth.\ A \vyp{580}{2007}{1485--1496.}

\bibitem{joaquinICRC}
J.~de Jes\'us \emph{et al.}, PoS \vyp{ICRC2023}{2023}{267.}

\bibitem{minos}
M.~Mathis  \emph{et al.}, J.\ Phys.\ Conf.\ Ser.\ \vyp{404.1 }{2012}{012--039.}


\end{thebibliography}
\end{document}